# Temperature dependent threshold for amplified emission from hybrid lead perovskite films

Yang Liu(刘洋)[1], Ju Wang(王雎)[1], Ning Zhu(朱宁)[1], Wei Liu(刘伟)[1], Cuncun Wu(吴存存)[1], Congyue Liu(刘聪越)[1], Lixin Xiao(肖立新)[1], Zhijian Chen(陈志坚)[1], Shufeng Wang(王树峰)[1,2*]

1 State Key Laboratory for Artificial Microstructure and Mesoscopic Physics, Department of Physics, Peking University, Beijing 100871, China.

2 Collaborative Innovation Center of Extreme Optics, Shanxi University, Taiyuan, Shanxi 030006, China.

**Abstract** The optical amplification emission of hybrid lead perovskite attracted great research interests. We systematically examined and compared temperature dependent optical amplification behavior of a series of organic-inorganic hybrid perovskite films of $(MA/FA)Pb(Br/I)_3$. The optical amplification threshold of the films showed considerable exponential increase towards the temperature increasing. We figured out that the critical temperature for the four films presented a sequence of FA+I < MA+I < FA+Br < MA+Br. Our systematical study is crucial for in depth understanding the fundamental mechanism of amplified emission of hybrid perovskite materials.

---

Hybrid lead perovskites were recently found exhibiting extraordinary optical amplification properties, which was regarded as optically pumped amplified spontaneous emission (ASE). It has been demonstrated in a variety of sample morphologies such as polycrystalline thin films,[1-8] nanocrystals,[9-12] nanowires,[13, 14] microcavity,[15, 16] microplates,[17-19] quantum dot,[20, 21] and distributed feedback cavity.[21, 22] The balanced charge transport and high gain characteristics of these hybrid perovskite materials make them the promising materials as optical gain media. These materials are also with many excellent properties like low-cost fabrication processes,[23, 24] strong light absorption, efficient photoluminescence,[25, 26] and long carrier lifetimes and diffusion lengths[27] have had numerous applications in optoelectronic devices beyond solar cells, including light-emitting diodes,[28] lasers,[29] and photodetectors.[30]

The high applicative interest of optical amplification emission phenomena from the solution-processed perovskites thin films stimulated the need for a systematical and clear understanding. The researches on its fundamental mechanism is still limited.[5, 7, 31] E.g., multiple phases with distinct optoelectronic properties existed in hybrid perovskites make it complicated to understand.[31-33] To understand the critical factors of optical amplification emission phenomena, the investigation of the optical properties as a function of the temperature is a very powerful tool. The systematic study for a series of hybrid perovskite thin films with its temperature related behavior is missing and ask for investigation to understand their behaviors.

In this paper, we report the systematic investigation of the temperature dependent ASE of the optical amplification from the films of four core hybrid lead perovskites, $MAPbI_3$, $FAPbI_3$, $MAPbBr_3$, and $MAPbBr_3$. Our study providing new view to the issue and fundamental support for application of amplified emission.

The perovskite films were deposited with flash evaporation technology, which applied a vacuum chamber was utilized for vacuum-assisted annealing of the samples after spin coating the mixture of perovskite precursor solution onto a glass substrate.[34] The films were annealed on a hot plate at 100℃ for 20 min. The morphology and crystallinity of the perovskite films were inspected by SEM (Fig. 1a-d). The as-formed perovskite films has full surface coverage on the substrates, with a grain size of about 300 nm −10μm.

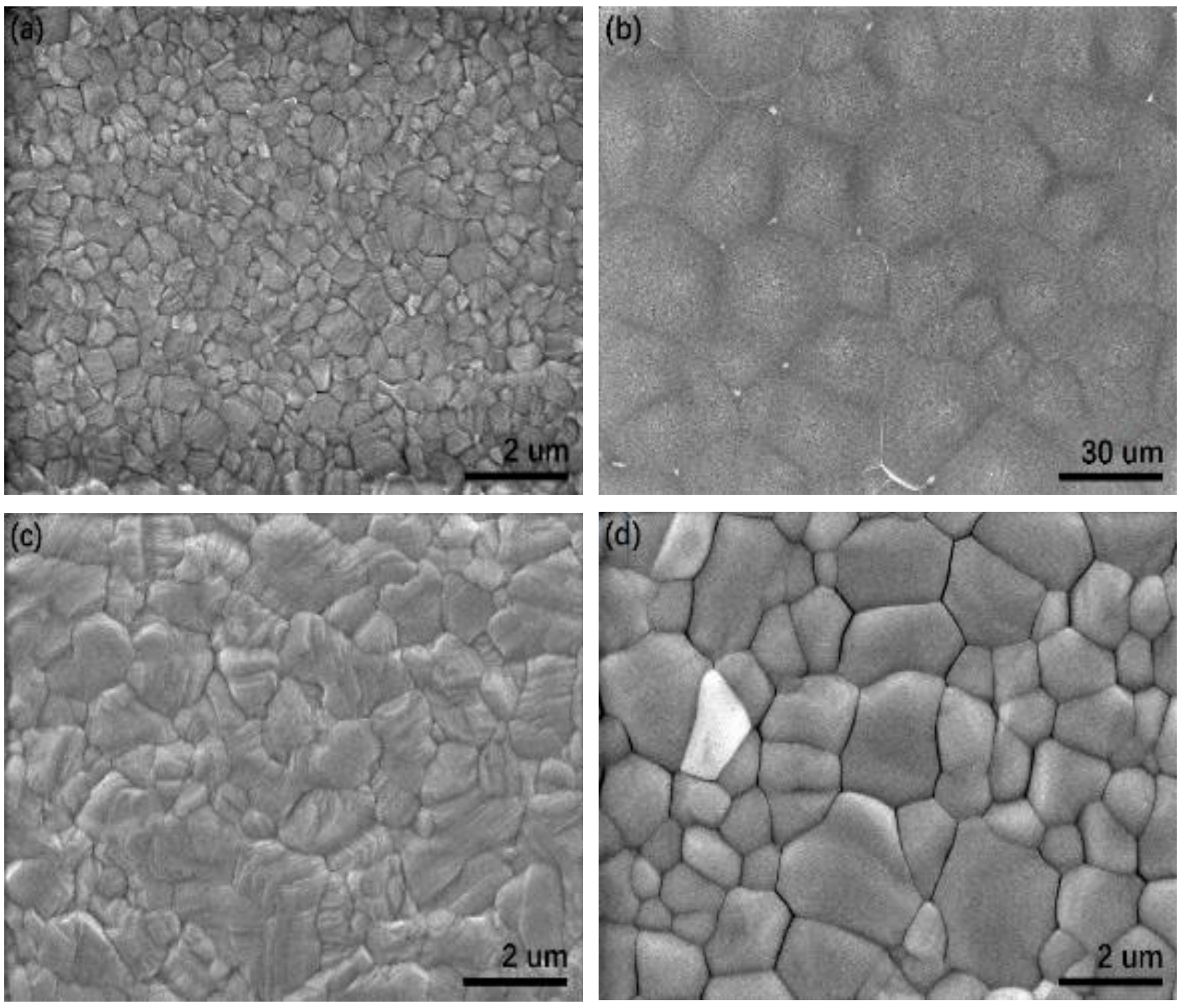

**Fig. 1** Scanning electron micrograph (SEM) of the top surfaces of (a)$MAPbI_3$, (b)$MAPbBr_3$, (c)$FAPbI_3$ and (d)$FAPbBr_3$ films.

The samples were mounted in a cryostat and cooled by feeding with liquid nitrogen. Femtosecond excitation pulses at 400nm, with frequency doubled from a Ti:sapphire regenerative amplifier that operates at 800 nm with 30 fs pulses with a repetition rate of 1 kHz (Coherent), were employed to excite the fluorescence. Since the lifetime is dependent to the excitation pulse energy, we applied time-resolved photoluminescent spectra through a streak camera system (Hamamatsu, ~20 ps resolution). Only the initial spectra at time zero are collected for further analysis.

The emission spectra at different temperature and excitation densities, were used to determine the temperature-dependent ASE in our samples. Fig. 2 shows the emission spectra from the four thin films at room temperatures (300 K). Taking Fig. 2(a) as example ($MAPbI_3$), When $P_{exc}$ was < 4.0 μJ/cm$^2$, the emission spectra showed similar broad peaks centered at ~760 nm with FWHM ~30 nm. When $P_{exc}$ was > 5.0 uJ/cm$^2$, a sharp emission band at 799.5 nm emerged from the red wing of the broad fluorescent spectra. The intensity of the new peaks at 799.5 nm grew fast towards the increment of pump energy, while the broadband fluorescent emission at 760 nm remained almost identical, when $P_{exc}$ increased from 5.0 to 7.9 uJ/cm$^2$. Similar behavior can be observed from $MAPbBr_3$, $FAPbI_3$, and $FAPbBr_3$ films. One noticeable difference for the $FAPbBr_3$ film to other samples is that its narrow peaks has much less red shift towards the broadband fluorescent peaks at ~552nm. The optical amplification features obtained from $MAPbI_3$, $MAPbBr_3$, $FAPbI_3$, and $FAPbBr_3$. Such behavior had been observed in other studies without systematic analysis. [4, 5, 8, 12] The temperature and pump intensity dependent peak shift can be found in supporting information (SI)

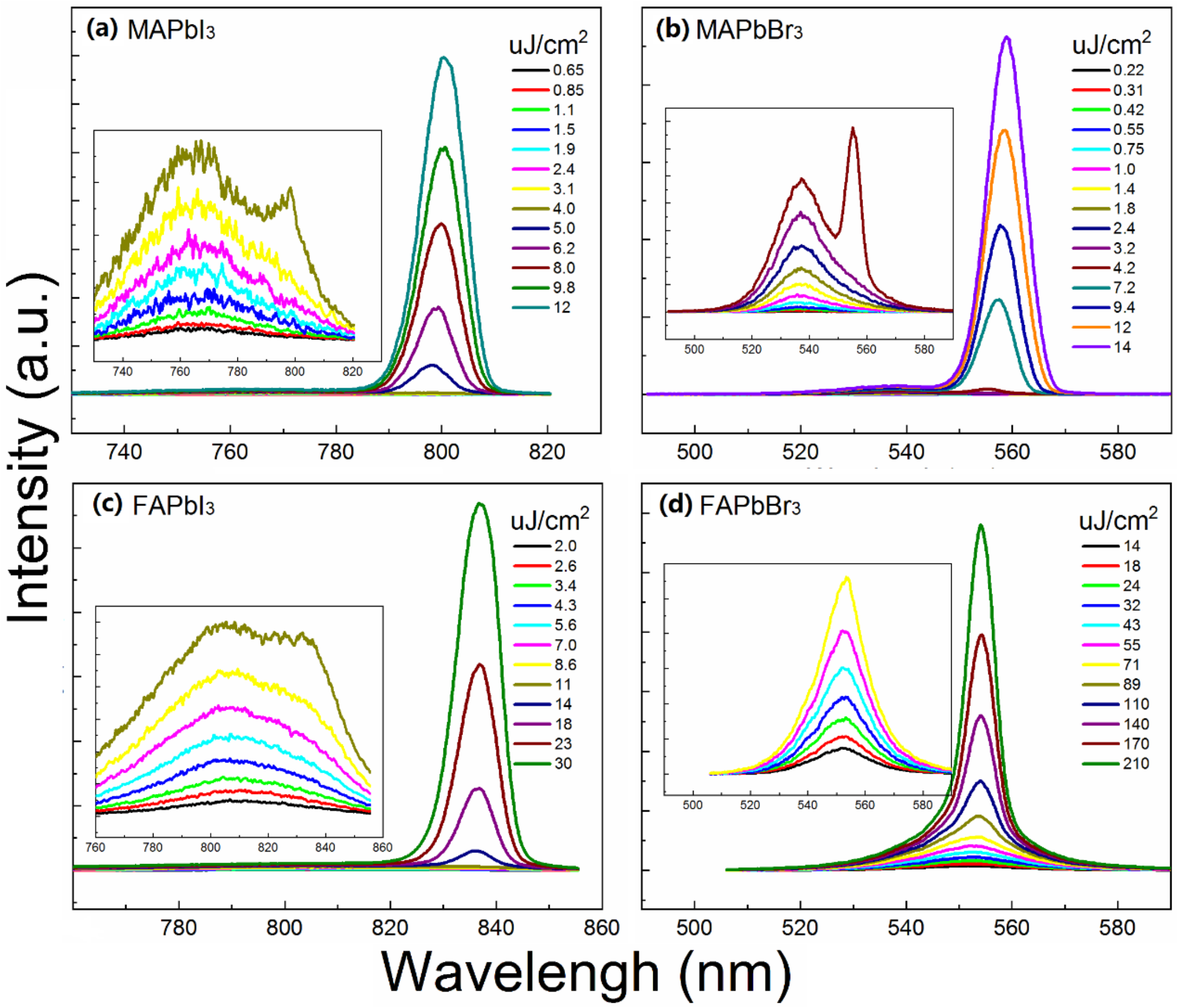

**Fig. 2** The distinct excitation density dependent emission spectra from perovskite thin films based on (a)$MAPbI_3$, (b)$MAPbBr_3$, (c)$FAPbI_3$, and (d)$FAPbBr_3$ at room temperatures (300 K).

The peak intensities increased towards the excitation density shows up a clear slope variation due to the appearance of ASE (see Fig. 3). The temperature dependent ASE threshold of the four films were estimated by fitting the ASE increment. The optical amplification threshold at room temperature (Fig. 4). As the temperature increases, we observe that the optical amplification threshold shows a curved continuous increase.

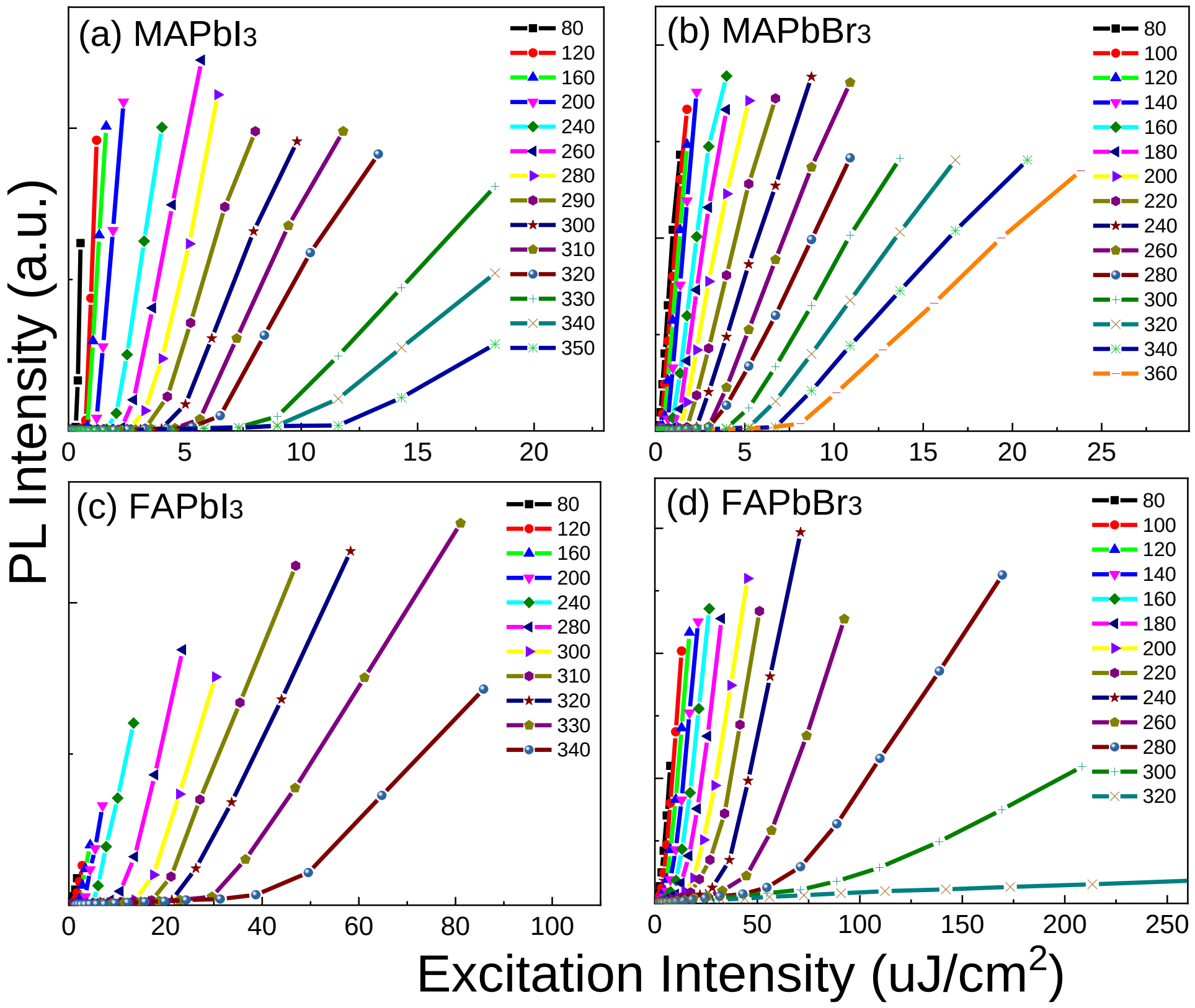

**Fig. 3** The total intensity dependence of (a)$MAPbI_3$, (b)$FAPbI_3$, (c)$MAPbBr_3$ and (d)$FAPbBr_3$ films on the excitation density.

The exponential temperature dependence of the current threshold in semiconductor diode lasers is empirically well-known as the following formula: [35]

$$P_{\mathrm{th}} = P_0 \cdot \exp\left(\frac{T}{T_0}\right)$$

Where $T$ is the temperature, $T_0$ is the characteristic temperature, and $P_0$ is the threshold when $T$ approaches zero. It is concluded in considering the exponential temperature dependent threshold resulting from the gain parameter and internal loss variations. The value of $T_0$ empirically represent the temperature-dependent sensitivity of threshold taking into account various factors. As seen in Fig. 4a-d, there are good agreement between the measurements and the fit, yielding various characteristic temperatures of $MAPbI_3$ ($T_0 = 56$K), $FAPbI_3$ ($T_0 = 42$K), $MAPbBr_3$ ($T_0 = 97$K), and $FAPbBr_3$ ($T_0 = 68$K) films respectively. For traditional semiconductor diode lasers, the values of $T_0$ are usually larger (~150-180K).[35] The measured values of $T_0$ in the 40–100 K range always mean significant change of the amplification threshold between 300 and 400K, the usual operating temperature range of lasers. By comparing these results, we could suggest that the perovskite with iodine has lower characteristic temperature to the ones with bromine, while the $FA^+$ brings the lower characteristic temperature towards the ones with $MA^+$. These difference indicate various sensitivities of the four perovskite to the temperature.

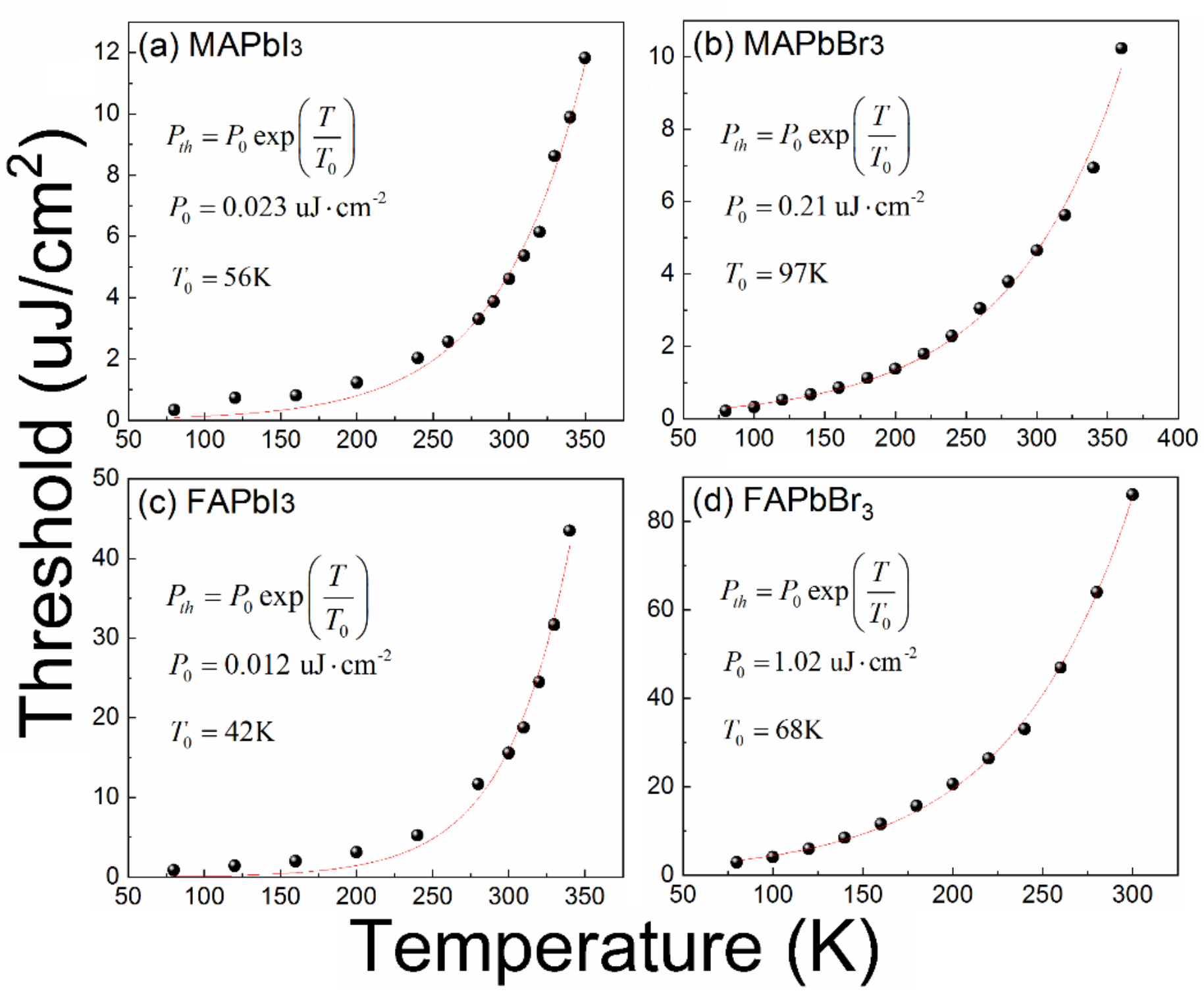

**Fig. 4** The temperature dependent amplification threshold of (a)$MAPbI_3$, (b)$MAPbBr_3$, (c)$FAPbI_3$ and (d)$FAPbBr_3$ films.

In conclusion, we systematically investigate the temperature dependent ASE threshold of four core organic-inorganic lead perovskite. Our results suggest that the ASE thresholds of perovskite thin films show exponential increase as the temperature increases. The exponential temperature dependence of the threshold yield various characteristic temperatures of $MAPbI_3$ ($T_0$ = 56K), $FAPbI_3$ ($T_0$ = 42K), $MAPbBr_3$ ($T_0$ = 97K) and $FAPbBr_3$ ($T_0$ = 68K) films respectively. Our results provide useful insights into the optical properties of hybrid perovskites, promoting their future applications in optoelectronic devices.

# Temperature dependent threshold for amplified emission from hybrid lead perovskite films

Yang Liu(刘洋)[1], Ju Wang(王雎)[1], Ning Zhu(朱宁)[1], Wei Liu(刘伟)[1], Cuncun Wu(吴存存)[1], Congyue Liu(刘聪越)[1], Lixin Xiao(肖立新)[1], Zhijian Chen(陈志坚)[1], Shufeng Wang(王树峰)[1,2]

1 State Key Laboratory for Artificial Microstructure and Mesoscopic Physics, Department of Physics, Peking University, Beijing 100871, China.

2 Collaborative Innovation Center of Extreme Optics, Shanxi University, Taiyuan, Shanxi 030006, China.

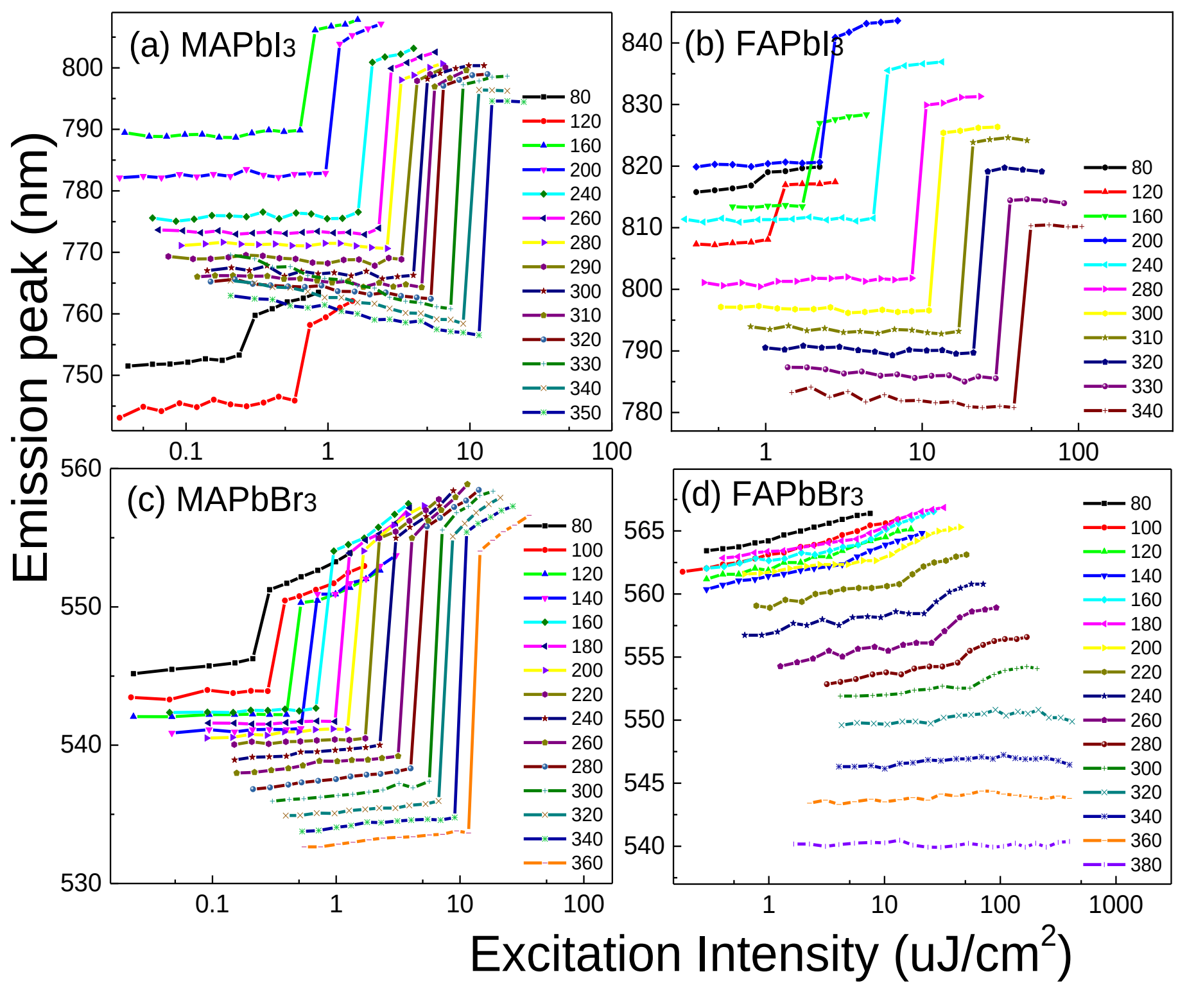

**Fig. S1** The temperature and excitation density dependent emission peak of (a)$MAPbI_3$, (b)$FAPbI_3$, (c)$MAPbBr_3$ and (d)$FAPbBr_3$ films.